\documentclass[aps,prd,nofootinbib,twocolumn,superscriptaddress,preprintnumbers]{revtex4-1}

\usepackage{amssymb}
\usepackage{amsmath}
\usepackage{epsfig}
\usepackage{hyperref}
\usepackage{breakurl}
\usepackage{bm}
\usepackage{color}
\usepackage{tikz}
\usetikzlibrary{decorations.markings}

\usepackage{subfigure}

\makeatletter
\def\simgt{\mathrel{\lower2.5pt\vbox{\lineskip=0pt\baselineskip=0pt
           \hbox{$>$}\hbox{$\sim$}}}}
\def\simlt{\mathrel{\lower2.5pt\vbox{\lineskip=0pt\baselineskip=0pt
           \hbox{$<$}\hbox{$\sim$}}}}
\makeatother

\newcommand{\be}{\begin{equation}}
\newcommand{\ee}{\end{equation}}

\newcommand{\bra}[1]{\langle #1 |}
\newcommand{\ket}[1]{| #1 \rangle}

\newcommand{\nn}{\nonumber}

\newcommand{\beq}{\begin{equation}}
\newcommand{\eeq}{\end{equation}}
\newcommand{\bea}{\begin{eqnarray}}
\newcommand{\eea}{\end{eqnarray}}

\def\topbotatom#1{\hbox{\hbox to 0pt{$#1\bot$\hss}$#1\top$}}

\begin{document}

\title{Breakdown of the Naive Parton Model in Super-Weak Scale Collisions}

\author{Matthew Baumgart}
\affiliation{Department of Physics, Arizona State University, Tempe, AZ 85287, USA}
\author{Ozan Erdo\u{g}an}
\affiliation{Department of Physics, Arizona State University, Tempe, AZ 85287, USA}
\author{Ira Z. Rothstein}
\affiliation{Department of Physics, Carnegie Mellon University, Pittsburgh, PA 15213, USA} 
\author{Varun Vaidya}
 \affiliation{Theoretical Division, MS B283, Los Alamos National Laboratory, Los Alamos, NM  87545, USA}

\begin{abstract}

In this letter we show that for observables which involve the measurement of weak charge in final states in hadronic collisions,  the standard parton model picture breaks down  at scales well above the weak scale due to non-factorizable electroweak corrections at leading order in the power expansion.
This implies that the resummation of these factorization-violating logarithms, which start at order  $\alpha_s^2\, \alpha_W^2 \log^4(Q^2/M_W^2)$, cannot be accomplished solely by following standard DGLAP evolution equations; other techniques will be needed to systematically sum large logarithms.

\end{abstract}

\maketitle

\section{Introduction}
It is a remarkable fact that one can make a prediction, within a systematic expansion, for a class of scattering cross sections between hadrons undergoing hard collisions. The proton represents a strongly coupled system composed 
of highly virtual entangled constituents, and yet when probed at short distances all of these complications
can be neatly packaged, for sufficiently inclusive observables, into a  parton distribution function (PDF), $f_{i/P}(\xi,\mu)$, which depends upon
the light-cone momentum fraction $\xi$ carried by the parton and renormalization (or factorization) scale $\mu$\@.

These PDFs determine the probability to find a parton of type $i$ within the proton with a given momentum fraction $\xi$.
Many hadronic cross sections can then be predicted in terms of a convolution of these PDFs 
with a quark-level scattering cross section.  For instance, a standard hadronic process is Drell-Yan (DY), $PP \rightarrow l^+ l^- + X$, the cross section for which can be written as a factorized convolution \cite{Bodwin:1984hc,CSS},
 \beq
 \label{eq:fact}
\sigma_{HH'}=\sum_{ij}f_{i/H}(\mu) \otimes  f_{j/H'}(\mu) \otimes \sigma_{i+j \rightarrow l^+l^-+X}(\mu) \ ,
\eeq
where $\sigma_{i+j \rightarrow l^+l^-+X}$ is the inclusive cross section for the partonic reaction, in which partons $i$ and $j$ produce a lepton pair and a hadronic state $X$, while 
$f_{i/H}$ and $f_{j/H'}$ are the PDFs (for partons of type $i$ and $j$, respectively) for the incoming anti-collinear hadrons. This result  is valid to leading order in an expansion in $\Lambda_{QCD}/Q$, where $\Lambda_{QCD}\sim 1$~GeV is the QCD
scale and $Q$ is the scale of the hard scattering. The long distance physics is  contained in $f_{i/H}$ and $ f_{j/H'}$ and is in general incalculable analytically, due to the confining nature of QCD.
We have made the renormalization group (RG) scale ($\mu$) dependence in this formula explicit because it reminds us that for our predictions to be accurate (systematic in any perturbative expansion) we must
sum logs of the ratio of scales in the theory (in this case $\log(Q^2/\Lambda_{QCD}^2)$). To resum these logs one  solves the
renormalization group equation (in this context called the DGLAP equation \cite{Altarelli:1977zs}) so that each function in the convolutions sits at its ``natural scale," so we choose, $\sigma(\mu=Q)$ and $f_{i/H}(\xi,\mu=\Lambda_{QCD})$.
This RG component of the parton model  can be ruined when the factorized form (\ref{eq:fact}) is violated for certain observables, as we now discuss.

\section{Glaubers and Factorization Violation in QCD}

 Equation~(\ref{eq:fact}) is what is known as a factorization theorem and can be elegantly described within the confines  of effective field theory (EFT), namely by soft-collinear effective theory (SCET), 
 where one splits the fields into a set of modes, each with its own unique momentum region \cite{SCET}. In the EFT, the stress energy tensor of the
leading order action can be written as a sum over the relevant modes,
\beq
\label{eq:T}
T^{\mu \nu}_{EFT}= T^{\mu \nu}_{n}+T^{\mu \nu}_{\bar n}+T^{\mu \nu}_{s} \ . 
\eeq
Here $n (\bar n)$ stands for ``(anti-)collinear" modes with light-cone momenta scaling as $k_{n}^\mu\sim(1, \lambda^2, \lambda)$, $k_{\bar n}^\mu\sim( \lambda^2,1 ,\lambda)$, respectively,  while ``soft'' modes scale as $k_s^\mu \sim (\lambda,\lambda,\lambda)$. For this notation we have decomposed momenta into $(p^+, p^-, p^\perp)$.  $\lambda$ is the power counting parameter, defined as $\Lambda_{IR}/Q$, where $\Lambda_{IR}$ is the relevant low
scale in the theory, which depends upon the observable.  As previously noted, for DY, $\Lambda_{IR} \sim \Lambda_{QCD}$, as long
as the masses ($m$) of all the particles in the loops are order $\Lambda_{QCD}$.  The hard scattering of top or bottom quarks is treated by introducing the notion of the top/bottom quark content of the proton, that is, introducing  PDFs for these massive quarks. Because $m_{t,b}\gg \Lambda_{QCD}$ 
the massive quark content of the proton is produced in perturbation theory via the splitting of collinear partons, which
is only sensible at RG scales well above the mass of the quark.
Note that although these PDFs are produced
in perturbation theory,  their effects become order one when $\alpha_s \log(Q^2/m^2) \sim 1$.
For scales above $m$, the quarks are taken to be massless, and the relevant infrared (IR) scale becomes $m$ so
that $\lambda \sim m/Q$. This change in the power counting will be relevant once we discuss electroweak corrections.

This splitting of degrees of freedom in the EFT leads to a tensor product Hilbert space
\beq
 \mid\!\! \psi \rangle= \mid \!\!\psi \rangle_n \otimes\mid \psi \rangle_{\bar n} \otimes \mid \psi \rangle_s \ , 
\eeq
 which facilitates factorization proofs of Ref.~\cite{hard}. 
The PDFs, $f$, are matrix elements of (anti-)collinear operators, so it is manifest that the various modes are completely detached as necessitated by Eq.~(\ref{eq:T}).
The only interactions between various modes are in the hard scattering kernel $\sigma$,  where perturbation theory is valid due to asymptotic freedom.

\begin{figure}[t!]
\centering
	\includegraphics[width=0.4\columnwidth]{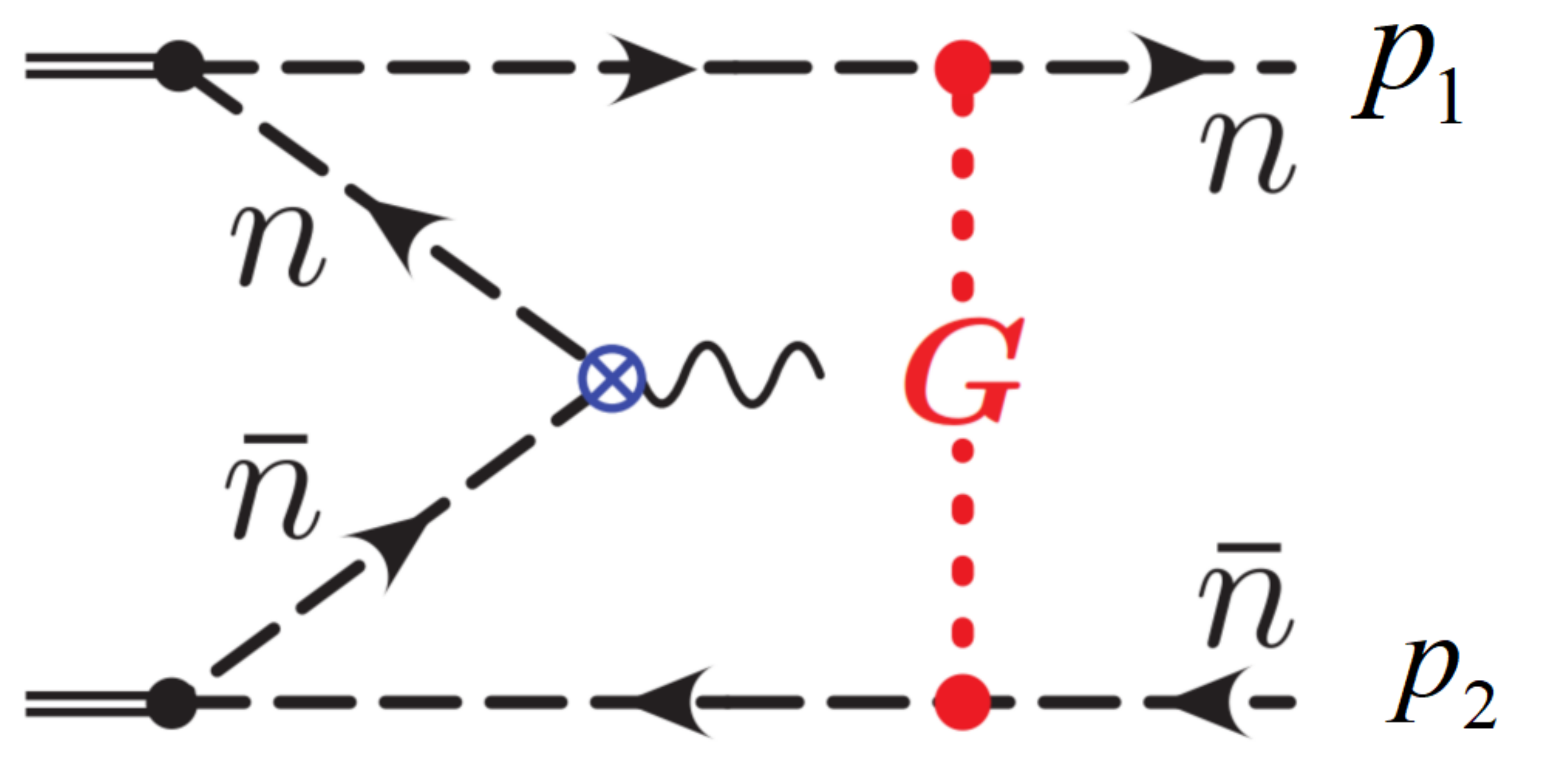} 
	\caption{Glauber exchange between collinear and anti-collinear spectators, which poses a threat to factorization.
		 The $G$ may include an arbitrary number of Glauber exchanges with box-like topology. The double line represents
		 the interpolating field for the hadron.
		} 
	\label{fig:SS}
\end{figure}

A natural question one should ask about Eq.~(\ref{eq:fact}) is why don't  interactions between spectator particles ({\it i.e.}~the infinite complement  to the active parton, which undergoes the hard scattering, within the proton) play a role in determining  the cross section? 
One would think that as long as soft gluons can be exchanged between spectator and  active partons, or between spectators
from each proton (as shown in Fig.~\ref{fig:SS}) the factorized form (\ref{eq:fact}) should be violated. Indeed, a simple one-loop calculation of the cross section shows
that in fact such (``Glauber gluon")  contributions do exist at leading power in $\Lambda/Q$. Yet, the result in Eq.~(\ref{eq:fact})  describes the
data remarkably well, a fact which seems to negate the existence of such soft exchanges.

The reason Eq.~(\ref{eq:fact}) properly describes DY is that in this process the Glauber contributions will cancel provided one freely integrates over the difference between the transverse momenta of the near-forward final state jets as proven in Ref.~\cite{CSS}\footnote{Recently a proof of the cancellation of Glaubers in double parton scattering was given in \cite{diehl}.}. 
 In SCET, the inclusion of Glauber gluons in hard scattering processes was introduced in \cite{glauber}. This work introduced  an operator formalism that can be used to treat Glaubers systematically in the power expansion, and rederived
 the aforementioned, sufficient condition for factorization violation, but did not prove the sufficient conditions (as done in \cite{CSS})
 for Glauber cancellation.

 When determining whether or not the Glaubers contribute to a given observable we must be sure that their effects are not already being accounted for in the canonical result (\ref{eq:fact})\footnote{Here we are following the EFT methodology. In \cite{CSS} the modes are
 not separated in terms of different fields.}. That is, Eq.~(\ref{eq:fact}) contains a set of Wilson lines
in the PDF, which account for some fraction of the long distance physics and we must not be fooled by (unphysical) Feynman diagrams into thinking that the Glaubers are not already contained in these Wilson lines.

 In SCET, it has been shown \cite{glauber} that the Glauber exchange between active and spectator quarks can be absorbed into the aforementioned Wilson lines, and thus are benign. However, it was
also shown that the same cannot be said for Glaubers exchanged between spectators. 
As mentioned above, for sufficiently 
inclusive observables, these Glauber gluons cancel in the final cross section. 
This cancellation occurs order by order, diagram by diagram,  after the sum  over all cuts of a diagram~\cite{CSS}.
If we consider the pure Glauber exchange  between spectators (as in Fig.~\ref{fig:SS}), to all orders, then the Drell-Yan amplitude is, in impact parameter ($b$) space,  proportional to a phase~\cite{glauber}
\beq
M \sim \int d^2b_\perp e^{-i \Delta p_\perp \cdot b_\perp} E(b_\perp, q_\perp) e^{i \phi(b_\perp)} \ , 
\eeq
with $\Delta p_\perp= (p_{2\perp}-p_{1\perp})/2 $ and $q_\perp= -(p_{1\perp}+p_{2\perp})/2$. 
The phase, 
\bea\label{eq:phi}
 \phi(b_\perp) 
  &=&  -{\rm\bf T}^A_1 \otimes {\rm\bf T}^A_2\, g^2(\mu)\, \frac{ \Gamma(-\epsilon) }{4\pi}
  \bigg( \frac{\mu|\vec b_\perp|e^{\gamma_E}}{2} \bigg)^{2\epsilon} \ , 
\eea
is a matrix in color space with ${\rm\bf T}^A_1$ and ${\rm\bf T}^A_2$ being the color matrix generators that commute with each other, and act on particle $1$ and $2$, respectively. The $\Gamma(-\epsilon)$ infrared divergence is akin to the usual Coulomb phase. It is noted that the phase in the squared amplitude vanishes only, as previously noted, if we perform the integral over $\Delta p_\perp$
over the full range of integration.

 An example of an observable where the Glaubers are not expected to cancel is beam thrust, where
one weights the final state transverse momentum distribution \cite{BT}, and we should not expect to (and cannot) fit the data using  the canonical formalism that leads to (\ref{eq:fact}).  We will use the term ``Kinematics Factorization Violation" (KFV)  when  factorization is violated by restricting the kinematics, to differentiate this scenario from the new one, which will be introduced below.

\section{Electroweak Glaubers}
Suppose we wish to calculate electroweak (EW) corrections to the DY process. While at $Q$  below the weak scale,
EW contributions are  suppressed by powers of $\Lambda^2_{IR}/M^2_W$,  for  $Q > M_W$ they become relevant especially given the fact that
electroweak corrections generate double logs of the form $\alpha_W \log^2(Q^2/M_W^2)$, 
as discussed further below. Thus, there has been considerable effort \cite{aneeshetal,baueretal} to resum EW corrections for LHC and future collider
predictions.  
The  question arises as to whether or not we  should expect that, by including PDFs for
all (massive) standard model particles, the parton model paradigm will still hold.  That is, can we calculate EW contributions
using Eq.~(\ref{eq:fact}) for arbitrarily large $Q$?

When addressing this question we begin by noting that the weak gauge group is not in the confining phase\footnote{The lack of clear distinction between Higgs/Coulomb and confining phase is of no relevance here.}. The significance of this  distinction lies
in the fact that we observe final state EW charge.
Indeed, DY in this sense is an exclusive observable since we measure leptons, which are now ``active partons"  in the final state.
 Furthermore, protons are charged in the weak sector and, as such, even completely inclusive quantities will be sensitive to the gauge-boson mass as the KLN theorem is violated \cite{Ciafaloni:1998xg,Ciafaloni2}, and as a consequence the cross sections will  include the aforementioned (Sudakov) double logarithms. 

To study electroweak Glauber effects, we will continue with our canonical example of Drell-Yan.
Above the weak scale, we allow for PDFs for all of the Standard Model (SM) particles, and the relevant IR scale becomes $M_W$ as in the case of massive quarks 
discussed above. 
In addition,  we must generalize our factorization theorem (\ref{eq:fact}) to account for the fact that the proton
is not an electroweak singlet, which again enhances the sensitivity to the IR via the introduction of Sudakov double logarithms of
the masses. As a result, the factorization theorem changes in two important ways. Firstly, the electroweak charge of the proton
implies that the bilinear operators which compose the PDF need no longer be a SM gauge singlet. 

Recall the QCD PDF that takes on the general form 
\beq
f^{(\mathrm{QCD})}_{i/P}(\xi)\!=\!\langle p \!\mid \! \int\! dy^- e^{i\xi p^+ y^-} O^\dagger_i(y^-) \Gamma W[y^-,0]O_i(0)\! \mid \! p \rangle \ , 
\label{PDF}
\eeq 
where $O$ is any (QCD) parton field and $W$ is the aforementioned Wilson line, which renders the bilinear gauge invariant, and 
$\Gamma$ is a spin tensor.  The generalization of this form to the full standard model gauge group must allow
for  bilinears that are not $SU(2)$ singlets. For instance, in addition to singlet PDFs we allow for a triplet of the form
 \beq\begin{split}
f^{(\mathrm{SM})a}_{i/P}(\xi)= 
\langle p \mid \!\! \int \!\! dy^- & e^{i\xi  p^+ y^-} \tilde O^\dagger_i(y^-) \tilde W[y^-,\infty]  \\
&\times \Gamma \,\tau^a\,\tilde W^\dagger[\infty,0]\tilde O_i(0) \!\! \mid p \rangle \ , 
\end{split}
\eeq
where $\tilde W$ are electroweak Wilson lines and the QCD Wilson line has been suppressed. Furthermore,
 the operators get modified
such that
\beq
\tilde O = S O \ , 
\eeq
where $S$ is a Wilson line composed  of soft gauge fields, which cancel in the case of singlet operators\footnote{This cancellation
occurs as a consequence of the need to multipole expand the soft fields \cite{GR} so that the Wilson lines cancel as a consequence of unitarity $S^\dagger S=1$.}.
The existence of the soft Wilson lines changes the way the logarithms are resummed in that the EW sector now
contains rapidity logarithms \cite{RRG}.
Secondly, since in DY we identify the leptons in the final state, and we are interested in resumming electroweak logs, 
we must account for final state cascades. The measurement of the final state isospin implies that we must include 
fragmentation functions $F_{i/j}$ in the convolution of Eq.~(\ref{eq:fact}), which give the probability of finding a given parton type $j$ in the
jet created by a mother parton of type $i$.  As in the case of the PDF, these fragmentation functions must also
be generalized to allow for non-singlet operators \cite{Baumgart:2014vma,Baumgart:2017nsr,Bauer:2018xag,Manohar:2018kfx}.

While we have had to modify Eq.~(\ref{eq:fact}), generalizing the factorization theorem to the standard model,  the notion of a PDF in the context of the parton 
model of hadrons  persists, in that the collinear partons in the two sectors are decoupled, as long as we can still
prove  that the Glauber contributions cancel.  In the next section, we will detail an example where they do not.

\section{Electroweak Factorization Violation} 

The measurement of final state isospin charge imperils the EW Glauber cancellation by its exclusive nature. 
We have noted that the QCD Glauber exchanges cancel upon summing over cuts.
However, given that we
are measuring the charge of the final state, it is clear that the cancellation can become ineffective since
some diagrams will not contribute for fixed-charge final states.
Another complication arises, {\it e.g.}~in DY,  
as the final states are now charged (SU(2)), and  we must consider
a new set Glauber exchanges, {\it i.e.}~those between spectators and active final states. An explicit calculation shows that Glauber integral is not sensitive to the collinear direction of the active parton. 
The underlying reason for this independence is the following identity
\beq
n^\prime \cdot p =  \frac{n^\prime \cdot n }{2} \bar n \cdot p  + \frac{ n^\prime \cdot \bar n }{2}  n \cdot p- n^\prime \cdot p_{\perp, n \bar n} =  \frac{n \cdot n^\prime }{2} \bar n \cdot p+ \mathcal{O}(\lambda) \ .
\eeq
Here $n/\bar n$ defines the direction of the spectator and $n^\prime$ is the direction of the final state active
parton. The residual $n^\prime$ dependence  will cancel due to reparametrization
invariance \cite{iain}.

\begin{figure}[b]
\centering
	\includegraphics[width=0.56\columnwidth]{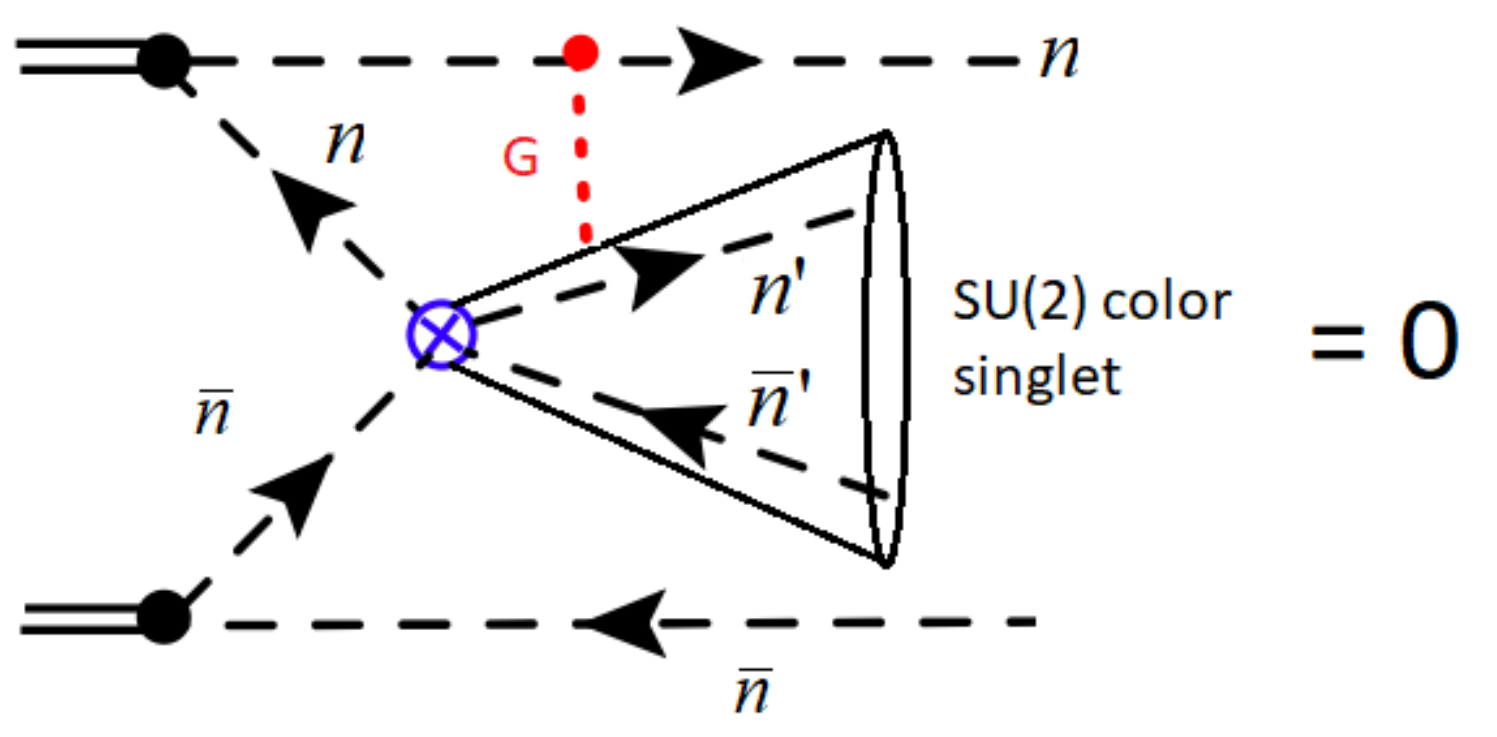} 
		\caption{Glauber exchange between spectators and active collinear partons in the final state. For a singlet current, the sum over Glauber graphs vanishes.		} 
	\label{fig:cs}
\end{figure}

\begin{figure}[t]
\centering
	\includegraphics[width=1\columnwidth]{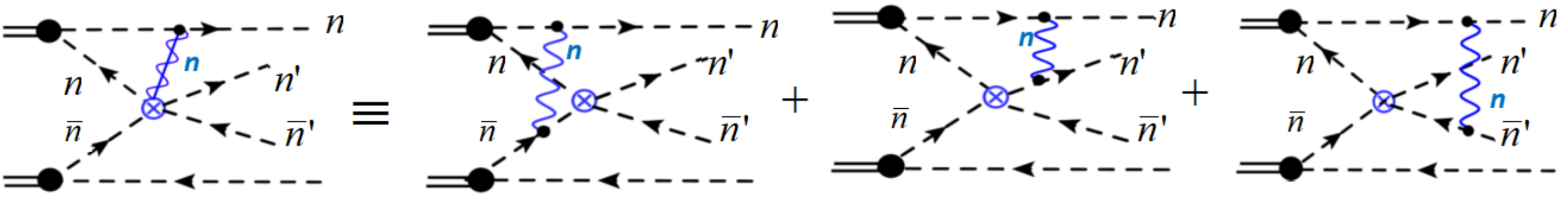} 
	\caption{Absorption of Glauber exchanges into collinear (EW) Wilson lines. A Wilson line for a particular collinear sector (here $\tilde W_n$) is built out of emissions from all the other well-separated collinear sectors.} 
	\label{fig:Wilson}
\end{figure}

The consequences of this independence are two-fold. If the current is a singlet then the
Glaubers will cancel explicitly in the sum over all possible
Glauber insertions, as shown in Fig.~\ref{fig:cs}, vanishes.
For a non-singlet current, as might be expected, this cancellation no longer occurs directly.
However, in this case, the final state active/spectator Glauber exchanges can be absorbed into collinear (EW) Wilson lines, which
now accompany the current. To see this, we note that the Wilson line for each collinear sector(say the n sector) is  in fact generated by summing
over n collinear emissions from all other collinear sectors ($n'$, $\bar n$, $\bar n'$). The  interactions between the $n$-spectator and all the other active partons (including the final state active partons)  are contained within the zero-bin of the corresponding interactions of the $W_n$ Wilson line and the $n$-spectator as illustrated by Fig.~\ref{fig:Wilson}, leaving the  factorization theorem unscathed\footnote{When the Glaubers are absorbed in this way, it fixes the direction of the Wilson line, i.e. whether it
stretches to $\pm \infty$ \cite{glauber}.}.

The generalization to the EW gauge group does not alter the fact that spectator-spectator
Glauber exchange persists. 
However, if we are measuring final state charge of the spectators, the cancellation is obstructed. For instance, suppose we are measuring top quark production rate which includes (the dominant) near forward direction. We should stress that this observable is not as inclusive as DY as we simply tag the presence of a $t$ or  $\bar t$ (or both) in the final state. Nevertheless, if we are also only differential in the kinematics of the produced leptons, we may expect the following straightforward modification to the standard PDF formulation for DY given in Eq.~\ref{eq:fact} to hold,  
\be
\frac{d \sigma}{d\Pi_{l^+l^-}} \Big |_{l^+ l^- + t \bar t + X} =\sum_{ij} B^{t}_i \otimes \bar B^{\bar t}_j\otimes \sigma_{i+j \rightarrow l^++l^-+X} ,
\label{eq:incorr}
\ee
where $\Pi_{l^+l^-}$ denotes whatever explicit measurement made on the leptons, and
\be\begin{split}
B^{t}_i(\xi) = \bra{p}\!\int\!\! dy^-&e^{i\xi p^+y^-}O^{\dagger}_i(y^-)W[y^-,0] \\
 & \times \Gamma\ket{t+X} \bra{t+X} O_i(0) \ket{p},
\end{split}\ee
while $t$ and $\bar t$ are replaced in $\bar B^{\bar t}_j$. This is essentially the same as the PDF defined in Eq.\ref{PDF} with the exception that the final state for this matrix element is no longer inclusive. If we, in fact, remove the requirement of top quarks in the final state, then we are once again fully inclusive over the spectators and recover Eq.~\ref{eq:fact}. 

\begin{figure}
\centering
\subfigure[]{\includegraphics[height=1.6cm]{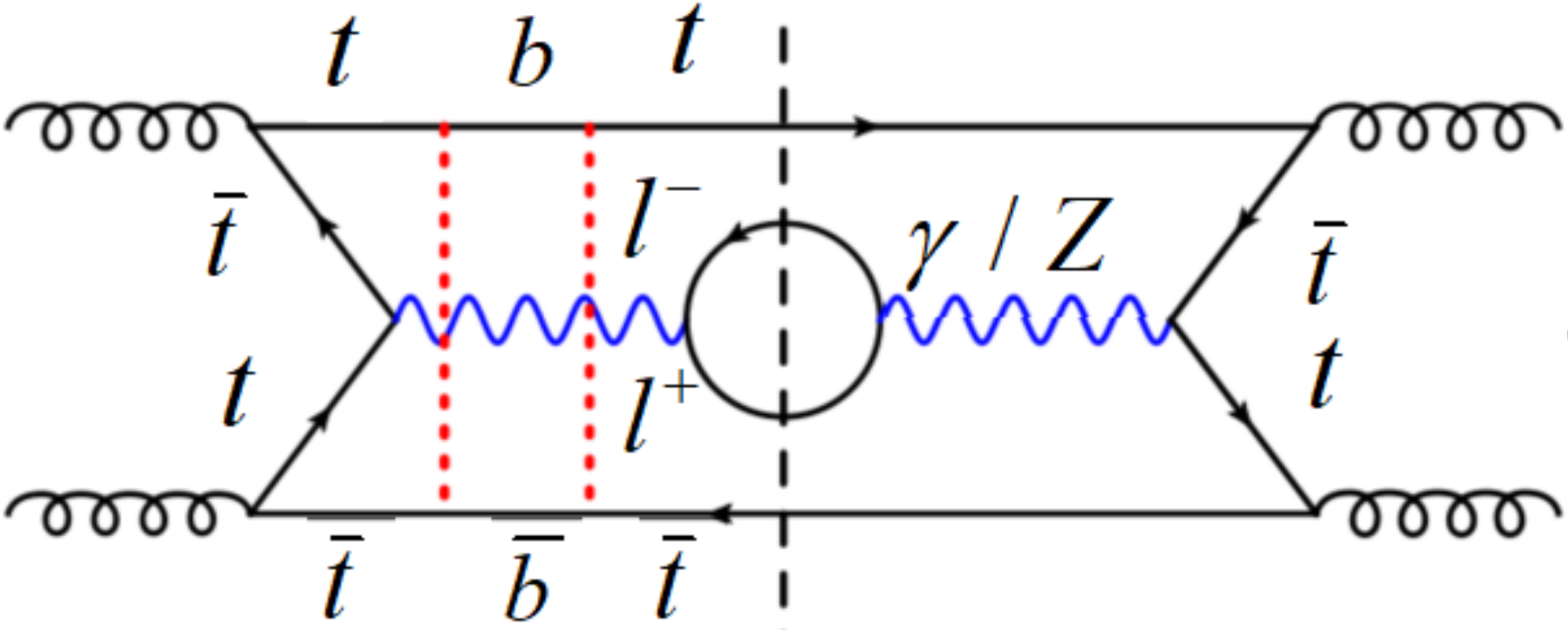}\label{fig:4a}}
\hspace{0.5cm}
\subfigure[]{\includegraphics[height=1.6cm]{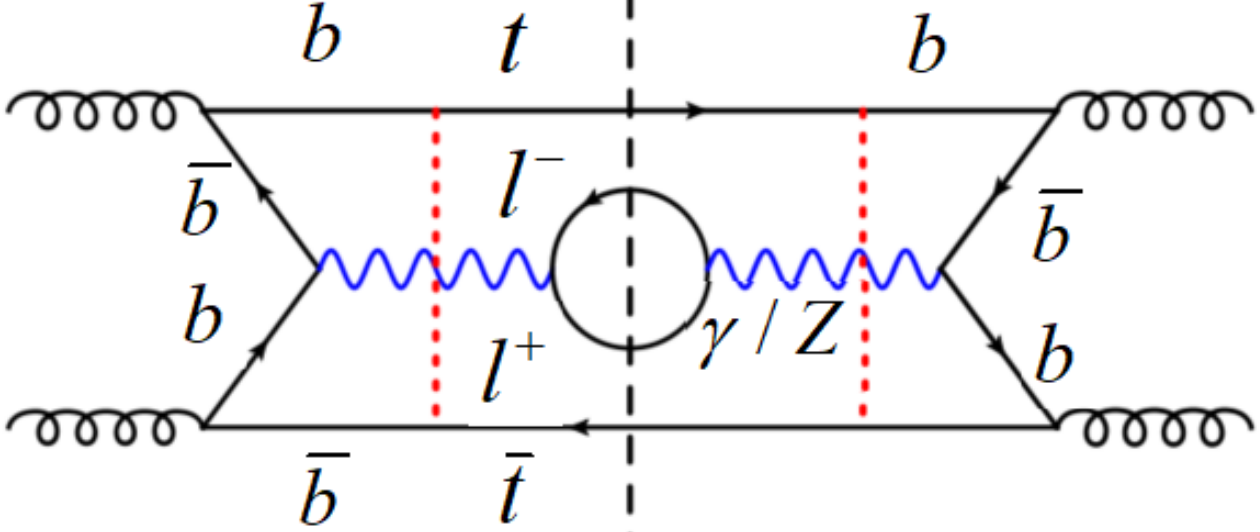}\label{fig:4b}}
\caption{Leading factorization violating diagrams involving EW Glauber exchanges.  The initial state arises from two gluons radiated with weak-scale offshellness from the protons.  The usual cancellation between these two graphs is blocked due to the
difference in isospin charges.} 
\label{figc:4loop}
\end{figure}

However, Eq.~\ref{eq:incorr} is {\it not correct}, as 
the
cancellation between the two diagrams shown in Fig.~\ref{figc:4loop} is blocked by the difference in the isospin charges which weight the diagrams. This prevents writing the observable as a simple convolution of two two-parton operators, like the $B_i$ above. Rather, the correct factorization theorem will require a new four-parton operator that involves active and spectator partons from both protons \cite{us}.
This type of ``Flavor Factorization Violation" (FFV) is distinct from KFV, and will arise for any observable where a particle
ID is made unless a cut is made enforcing large $p_\perp \gg M_W$.\footnote{Some authors prefer to reserve the term factorization violation to refer only to quantities participating at the hard vertex. We use it here though, to refer to the fact that radiation exchanged between different collinear sectors cannot be written as a Wilson line.}  
The contribution of these factorization violating diagrams is given by
\bea
\label{eq:expldiagram}
&& \mathrm{Im}\, M_{4a}\sim P_{qg}(x)P_{qg}(y)\!\!\int\!\! d^2p_\perp  d^2\bar p_\perp \frac{dl^2_\perp}{l_\perp^2+M_W^2} \frac{dl^{\prime 2}_\perp}{l^{\prime 2}_\perp+M_W^2} \nn \\ &&\times\frac{(\vec p_\perp - \vec l_\perp) \cdot  (\vec p_\perp - \vec l^\prime_\perp)}{(\vec p_\perp - \vec l_\perp)^2+m_t^2}\frac{(\vec {\bar p}_\perp + \vec l_\perp) \cdot  (\vec {\bar p}_\perp + \vec l^\prime_\perp)}{(\vec {\bar p}_\perp + \vec l_\perp)^2+m_t^2} \nn \\
&&\times\frac{1}{(\vec {\bar p}_\perp + \vec l^\prime_\perp)^2+m_t^2} \frac{1}{(\vec { p}_\perp - \vec l^\prime_\perp)^2+m_t^2}\, \delta(xy-Q^2/\hat s) \ ,
 \eea
where $P_{qg}$ is the quark-gluon splitting function and $x$, $y$ are the momentum fractions of the
incoming gluons. Performing the loop integrals we find
\beq
\label{LL}
\mathrm{Im}\, M_{4a} \sim \alpha_s^2 \alpha_W^2 \log^4(M_W^2/Q^2) \ , 
\eeq
where we have set the RG scale $\mu$ to the scale of hard scattering and $m_q \sim M_W$. Note that these logarithms cannot be resummed by the canonical running of the PDF, 
as expected, since the Glaubers talk between the two proton jets. This dimensionless correction factor in Eq.~\ref{LL} already reaches 0.1 for $Q \sim$ 10 TeV.  The detailed quantitative study of these contributions is beyond the scope of this work, but this basic estimate shows that the numerical importance of this effect need not be small.

For the case when all the hard interaction currents have no weak charge (e.g. $P+P \rightarrow$ jets), again assuming
we measure some weak charge in the final state, we expect a cancellation between diagrams with the topology of Figs.~\ref{fig:4a}~and~\ref{fig:4b}, which will eliminate the leading quartic logarithmic terms.
There will be mass dependent terms 
at leading order in the power expansion \cite{mass} since $m_t \sim M_W$
that lead to asymptotic logs of the form 
\beq
\alpha_s^2 \alpha_W^2 \left(M_t^2/M^2_W\right) \log(M_W^2/Q^2) \ .
\eeq
All the logarithms obtained here via the electroweak Glauber exchange are not associated with the BFKL \cite{BFKL} logarithms usually associated with near-forward scattering (rapidity logs). Rather these are renormalization group logarithms that are resummed
by augmenting the space of four parton operators \cite{us}. 
There will be BFKL running as well, that will be subleading to (\ref{LL}).

	\section{Conclusions}
We have shown that the parton model picture, where a systematic calculation of a hard scattering
cross section follows from a factorization theorem involving PDFs, no longer holds when
directly extended to the EW sector and applied to EW observables in proton scattering.  
In the EW case, a new mechanism for factorization violation arises when final state isospin  is measured for states with  ($y\gtrsim 1$, $p_\perp \gtrsim 100$~GeV), as the requisite cancellation is obstructed due to the dearth of allowed cuts. As a result, one may not simply add leptonic and EW gauge-boson PDFs (as put forward in \cite{baueretal}) and accurately sum all of the EW logarithms using DGLAP physics.

This violation, while numerically small at the weak scale, will become logarithmically enhanced for $Q\gg M_W$, and
will become relevant for a myriad of interesting observables, including multi-Higgs  production.  
Moreover, the factorization-violating logarithms are not summed by the BFKL equations nor by the DGLAP equation, but will need to be treated within a new formalism that will be presented elsewhere~\cite{us}.  Furthermore, we see that for certain processes, like forward top quark production, novel operators beyond the usual PDF framework are required.  From a field theoretic standpoint
this implies that, as we push the energy frontier beyond the weak scale, various other theoretical techniques, beyond the now-textbook DLGAP material, will need to be harnessed/developed, at least for a  class of relevant observables.

{\it Acknowledgments:} 
We would like to thank Duff Neill and Iain Stewart for useful discussions. MB and OE are supported by the U.S. Department of Energy, under grant number DE-SC-0000232627. The work of OE was also supported by the DOE grant No.~DE-SC-0010118. IZR is supported by DOE HEP grants DE-FG02-04ER41338 and FG02-06ER41449.  VV is supported by the U.S. Department of Energy through the Office of Nuclear Physics under Contract DE-AC52-06NA25396 and  through the LANL/LDRD Program. MB and OE would like to thank Los Alamos National Laboratory for their hospitality, while part of this work was done.


\begin{thebibliography}{99}


\bibitem{Bodwin:1984hc} 
  G.~T.~Bodwin,
  Phys.\ Rev.\ D {\bf 31}, 2616 (1985)
  Erratum: [Phys.\ Rev.\ D {\bf 34}, 3932 (1986)].

   \bibitem{CSS}
 J.~C.~Collins, D.~E.~Soper and G.~Sterman,
 Nucl.\ Phys.\ B {\bf 261}, 104 (1985); 
 Nucl.\ Phys.\ B 
 {\bf 308}, 833 (1988).



\bibitem{Altarelli:1977zs} 
  G.~Altarelli and G.~Parisi,
  Nucl.\ Phys.\ B {\bf 126}, 298 (1977);
V.~N.~Gribov and L.~N.~Lipatov,
  Sov.\ J.\ Nucl.\ Phys.\  {\bf 15}, 438, 675 (1972)
  [Yad.\ Fiz.\  {\bf 15}, 781, 1218 (1972)];
 Y.~L.~Dokshitzer,
  Sov.\ Phys.\ JETP {\bf 46}, 641 (1977)
  [Zh.\ Eksp.\ Teor.\ Fiz.\  {\bf 73}, 1216 (1977)]

\bibitem{SCET} 
 C.~W.~Bauer, S.~Fleming and M.~E.~Luke,
  Phys.\ Rev.\ D {\bf 63}, 014006 (2000)
  [hep-ph/0005275];

  
  C.~W.~Bauer, S.~Fleming, D.~Pirjol and I.~W.~Stewart,
  Phys.\ Rev.\ D {\bf 63}, 114020 (2001)
  [hep-ph/0011336];

  
  C.~W.~Bauer, D.~Pirjol and I.~W.~Stewart,
  Phys.\ Rev.\ D {\bf 65}, 054022 (2002)
  [hep-ph/0109045].

  
  \bibitem{hard} 
  C.~W.~Bauer, S.~Fleming, D.~Pirjol, I.~Z.~Rothstein and I.~W.~Stewart,
  Phys.\ Rev.\ D {\bf 66}, 014017 (2002)
  [hep-ph/0202088].

  


  \bibitem{diehl}
   M.~Diehl, J.~R.~Gaunt, D.~Ostermeier, P.~Pl\"{o}{\ss}l and A.~Sch\"{a}fer,
  JHEP {\bf 1601}, 076 (2016)
  [arXiv:1510.08696 [hep-ph]].


  \bibitem{glauber}
  I.~Z.~Rothstein and I.~W.~Stewart,
  JHEP {\bf 1608}, 025 (2016)
  [arXiv:1601.04695 [hep-ph]].


  \bibitem{BT} 
  S.~Alioli, C.~W.~Bauer, S.~Guns and F.~J.~Tackmann,
  Eur.\ Phys.\ J.\ C {\bf 76}, no. 11, 614 (2016)
  [arXiv:1605.07192 [hep-ph]].

  
 \bibitem{aneeshetal}
  J.~Y.~Chiu, A.~Fuhrer, R.~Kelley and A.~V.~Manohar,
  Phys.\ Rev.\ D {\bf 80}, 094013 (2009)
  [arXiv:0909.0012 [hep-ph]].
  
  \bibitem{baueretal}
    C.~W.~Bauer, N.~Ferland and B.~R.~Webber,
  JHEP {\bf 1708}, 036 (2017)
  [arXiv:1703.08562 [hep-ph]].
  

  
\bibitem{Ciafaloni:1998xg} 
  P.~Ciafaloni and D.~Comelli,
  Phys.\ Lett.\ B {\bf 446}, 278 (1999)
  [hep-ph/9809321].
  \bibitem{Ciafaloni2} 
  M.~Ciafaloni, P.~Ciafaloni and D.~Comelli,
  Phys.\ Rev.\ Lett.\  {\bf 84}, 4810 (2000)
  [hep-ph/0001142];
  Nucl.\ Phys.\ B {\bf 589}, 359 (2000)
  [hep-ph/0004071];
  Phys.\ Lett.\ B {\bf 501}, 216 (2001)
  [hep-ph/0007096];
  Phys.\ Rev.\ Lett.\  {\bf 88}, 102001 (2002)
  [hep-ph/0111109].
	
	
  
  \bibitem{GR}
  B.~Grinstein and I.~Z.~Rothstein,
  Phys.\ Rev.\ D {\bf 57}, 78 (1998)
  [hep-ph/9703298].
  
  \bibitem{RRG} 
  J.~Y.~Chiu, A.~Jain, D.~Neill and I.~Z.~Rothstein,
  JHEP {\bf 1205}, 084 (2012)
  [arXiv:1202.0814 [hep-ph]];
  J.~y.~Chiu, A.~Jain, D.~Neill and I.~Z.~Rothstein,
  Phys.\ Rev.\ Lett.\  {\bf 108}, 151601 (2012)
  [arXiv:1104.0881 [hep-ph]].
  
  
\bibitem{Baumgart:2014vma} 
  M.~Baumgart, I.~Z.~Rothstein and V.~Vaidya,
  Phys.\ Rev.\ Lett.\  {\bf 114}, 211301 (2015)
    [arXiv:1409.4415 [hep-ph]]. 

 \bibitem{Baumgart:2017nsr}
  M.~Baumgart, T.~Cohen, I.~Moult, N.~L.~Rodd, T.~R.~Slatyer, M.~P.~Solon, I.~W.~Stewart and V.~Vaidya,
  JHEP {\bf 1803}, 117 (2018)
  [arXiv:1712.07656 [hep-ph]].

  \bibitem{Bauer:2018xag} 
   C.~W.~Bauer, D.~Provasoli and B.~R.~Webber,
  arXiv:1806.10157 [hep-ph].
  
\bibitem{Manohar:2018kfx} 
  A.~V.~Manohar and W.~J.~Waalewijn,
  JHEP {\bf 1808}, 137 (2018)
  [arXiv:1802.08687 [hep-ph]].

 \bibitem{iain}
  A.~V.~Manohar, T.~Mehen, D.~Pirjol and I.~W.~Stewart,
  Phys.\ Lett.\ B {\bf 539}, 59 (2002)
  [hep-ph/0204229].

  \bibitem{us} M.~Baumgart, O.~Erdo\u{g}an, I.~Z.~Rothstein and V.~Vaidya, work in progress.

  \bibitem{mass}
  I.~Z.~Rothstein,
  Phys.\ Rev.\ D {\bf 70}, 054024 (2004)
    [hep-ph/0301240]; 
  A.~K.~Leibovich, Z.~Ligeti and M.~B.~Wise,
  Phys.\ Lett.\ B {\bf 564}, 231 (2003)
  [hep-ph/0303099].

  
 \bibitem{BFKL}  E.~A.~Kuraev, L.~N.~Lipatov and V.~S.~Fadin,
  Sov.\ Phys.\ JETP {\bf 45}, 199 (1977)
  [Zh.\ Eksp.\ Teor.\ Fiz.\  {\bf 72}, 377 (1977)]. I.~I.~Balitsky and L.~N.~Lipatov,
  Sov.\ J.\ Nucl.\ Phys.\  {\bf 28}, 822 (1978)
  [Yad.\ Fiz.\  {\bf 28}, 1597 (1978)].

 





\end{thebibliography}
\end{document}